\documentclass[pra,twocolumn,aps,superscriptaddress,showpacs,floatfix]{revtex4-1}
\usepackage{amssymb}
\usepackage{graphicx}
\usepackage{dcolumn}
\usepackage{bm}
\usepackage{amsmath}
\usepackage{amsbsy}
\usepackage{amsmath}
\usepackage{cases}
\usepackage{graphicx}
\usepackage{color}
\usepackage{epstopdf}
\usepackage{mathrsfs}
\usepackage{multirow}
\usepackage[colorlinks,linkcolor=magenta,citecolor=blue,urlcolor=blue]{hyperref}
\usepackage{etoolbox}
\usepackage{multirow}
\makeatletter
\patchcmd{\@makecaption}
  {\scshape}
  {}
  {}
  {}
\makeatother

\makeatletter
\newcommand{\Rmnum}[1]{\expandafter\@slowromancap\romannumeral #1@}
\makeatother

\begin{document}
\title{Reentrant Localization Transitions in a Topological Anderson Insulator: A Study of a Generalized Su-Schrieffer-Heeger Quasicrystal}
\author{Zhanpeng Lu}
\affiliation{Institute of Theoretical Physics and State Key Laboratory of Quantum Optics and Quantum Optics Devices, Shanxi University, Taiyuan 030006, China}
\author{Yunbo Zhang}
\email{ybzhang@zstu.edu.cn}
\affiliation{Key Laboratory of Optical Field Manipulation of Zhejiang Province and Physics Department of Zhejiang Sci-Tech University, Hangzhou 310018, China}
\author{Zhihao Xu}
\email{xuzhihao@sxu.edu.cn}
\affiliation{Institute of Theoretical Physics and State Key Laboratory of Quantum Optics and Quantum Optics Devices, Shanxi University, Taiyuan 030006, China}
\affiliation{Collaborative Innovation Center of Extreme Optics, Shanxi University, Taiyuan 030006, China}


\begin{abstract}
 We study the topology and localization properties of a generalized Su-Schrieffer-Heeger (SSH) model with a quasi-periodic modulated hopping. It is found that the interplay of off-diagonal quasi-periodic modulations can induce topological Anderson insulator (TAI) phases and reentrant topological Anderson insulator (RTAI), and the topological phase boundaries can be uncovered by the divergence of the localization length of the zero-energy mode. In contrast to the conventional case that the TAI regime emerges in a finite range with the increase of disorder, the TAI and RTAI are robust against arbitrary modulation amplitude for our system. Furthermore, we find that the TAI and RTAI can induce the emergence of reentrant localization transitions. Such an interesting connection between the reentrant localization transition and the TAI/RTAI can be detected from the wave-packet dynamics in cold atom systems by adopting the technique of momentum-lattice engineering.
\end{abstract}

\maketitle
\section{Introduction}
Anderson localization is a ubiquitous phenomenon in the condensed matter where the wave-like behavior of particles becomes localized in a disordered medium, preventing their propagation. This phenomenon was first introduced by P. W. Anderson in 1958 and has since been observed in a wide range of physical platforms \cite{Anderson1958,Ramakrishnan1985RMP,Billy2008Nat,Roati2008Nat,Chabanov2000Nat,Pradhan2000PRL,Lahini2009PRL}. For low-dimensional systems, the single-parameter scaling theory predicts that an arbitrarily small on-site random disorder induces all states to be localized \cite{Mott1987JPC}. However, such a scaling theory is invalid in a quasi-periodic system, for which low-dimensional quasicrystals may exhibit a metal-to-insulator transition. The paradigmatic example is the Aubry-Andr\'{e} (AA) model \cite{Aubry1980,Harper1955,Longhi2019PRL,Longhi2019PRB}, which has been experimentally realized in cold atoms \cite{Roati2008Nat} and photonic crystals \cite{Kraus2012PRL,Segev2013NP}. In some extended AA models, one can obtain an intermediate phase with coexisting extended and localized states separated by a critical energy called the mobility edge (ME) \cite{Biddle2010PRL,Ganeshan2015PRL,XJL2022,XJL2023,TL2023,WC2022,DP2023}. Moreover, quasi-periodic systems exhibit many unique properties, such as reentrant localization transition \cite{SRoy2021PRL}. It indicates the system can undergo two localization transitions. Subsequently, this phenomenon has also been observed in many other quasi-periodic models \cite{Padhan2022PRB,SRoy2022PRB,ZWZPRA2022,SA2023PRB,JXPPRB1,JXPPRB2,PSARIXV,LXZARIXV1,LXZARIXV2,CWNJP2021,XPJ2021CPB,LZhou2022PRB,WH2022PRB,HW2023PRB}.
This discovery deepens the understanding of the localization transition in quasi-periodic systems.

On the other hand, topological insulators, known for their quantized electronic transport in bulk states and the presence of nontrivial in-gap modes, have garnered significant attention over the past few decades.  Topology and disorder exhibit many fantastic connections, from the similarity of one-dimensional (1D) quasi-periodic and 2D Hofstadter lattices, to the connection between the random matrix and the classification of topological phases \cite{Nakajima2021NP,SC1,SC2}. One hallmark property of the topological insulators is the robustness of nontrivial edge states against weak disorder in the underlying lattice. This robustness arises because quantized transport along the edge states, which are protected against backscattering by topological invariants. However, when the disorder amplitude is large enough, the band gap closes, and the system typically transitions into a trivial phase \cite{Su1979,Thouless1982,Hasan2010,QXliangRMP2011,BansilRMP2016,Chiu2016,ArmitageRMP2018,Prodan2010,Song2019PRL,Zhu2019prl,XZhao2020PRA1,Bo2021,Cai2013,Potter2012,Pan2021}. Interestingly, disorder can also induce nontrivial topology in a system that initially exhibits trivial bands, a phenomenon known as the topological Anderson insulator (TAI). This phase is marked by the emergence of topologically protected edge modes and quantized topological charges \cite{Shen2009,Guo2010,Zhang2012}. The TAI phase was first predicted in 2D quantum wells with random disorder \cite{Shen2009} and has since been demonstrated across a variety of systems, including Su-Schrieffer-Heeger (SSH) chain \cite{DZhang2020}.
Subsequently, Groth et al. \cite{Groth2009} introduced an effective medium theory, the self-consistent Born approximation, to explain the mechanism behind TAI formation. During the transition from a trivial to a nontrivial phase, the topological mass changes sign from negative to positive with the zero value of the topological mass marking the critical point of the topological phase transition. Here, the topological mass refers to the self-energy correction induced by random disorder \cite{Groth2009,Girschik2013,DZhang2022,YPW2022,SNL2022}.
The TAI phase has been extensively studied in various theoretical models \cite{Song2012,Hughes2014,Zhang2019,Tangg2020,Borchmann2016,Hua2019,GQZ2021,Velury2021,WJZ2022,XXC2009} and observed experimentally in various artificial systems such as 2D photonics system \cite{Sttzer2018,Liu2020} and 1D engineering synthetic chiral symmetric wires \cite{Meier2018}. For disorder-induced TAI, it is generally observed that the bulk states remain localized. However, recent studies have shown that in quasi-periodic modulation-induced TAI phase, the bulk states exhibit unique localization properties \cite{Longhi2020,LZP2022,DZhang2022}.

This paper investigates a generalized SSH model with off-diagonal quasi-periodic modulations exhibiting exotic topological and localization phenomena. First, we numerically calculate the topological phase diagram by the real-space winding number to characterize the effects of off-diagonal quasi-periodic modulations on topology.
According to the topological phase diagram, we found that the off-diagonal quasi-periodic modulations can induce three different topological phases: the topological insulator phase, the reentrant topological Anderson insulator phase(RTAI), and the topological Anderson insulator phase(TAI). The RTAI and TAI phases in our system exhibit the robustness of the nontrivial topology under strong disorder strength condition. Such topological features are also characterized by the divergence of the zero mode's localization length. Furthermore, we study the localization properties of our system and verify the existence of the reentrant localization transition. By comparing the localization and topological phase diagrams, we find that the reentrant localization transition coincides with the RTAI and TAI phase transitions, and a physical explanation is given. Finally, we illustrate that the relationship between the TAI and the reentrant localization transition can be detected by the wave-packet dynamics in the momentum-lattice system.
\section{Model and Hamiltonian}
\begin{figure}[tbp]
	\includegraphics[width=0.5\textwidth]{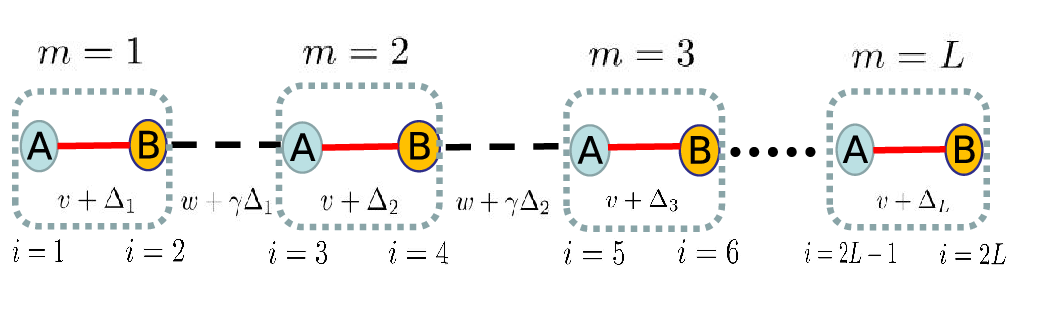}
	\caption{(Color online) Schematic illustration of the generalized quasi-periodic SSH model with two sublattices ($A$ and $B$) in each unit cell shown in the dotted box. This is a chain with $L$ unit cells, and the length of the chain is $2L$. Here, $m$ is the index of the unit cell, and $i$ represents the lattice site. $w+\gamma\Delta_m$ and $v+\Delta_m$ are the $m$-dependent intercell and intracell hopping amplitudes, respectively.}
	\label{Fig1}
\end{figure}

The reentrant localization transition is first observed in a SSH model with the staggered quasi-periodic on-site potential, which breaks the chiral symmetry of the SSH chain and is topologically trivial. To detect the relationship between the reentrant localization and the topological transition, we consider a generalized SSH model with quasi-periodic modulated hopping, which preserves the chiral symmetry and is depicted in Fig. \ref{Fig1}. The dimerized tight-binding model can be described by
\begin{equation}
\small
	H=\sum_{m=1}^Lv_m( c^{\dag}_{m,B}c_{m,A}+H.c.)+\sum_{m=1}^{L-1}w_m( c^{\dag}_{m+1,A}c_{m,B}+H.c.).
	\label{eq1}
\end{equation}
This is a chain of $L$ unit cells consisting of two sublattices labeled by $A$ and $B$, and the length of the lattice is $2L$. As shown in Fig. \ref{Fig1}, $m$ is the index of the unit cell, and $i$ represents the lattice site. $c^{\dag}_{m,A}$ ($c^{\dagger}_{m,B}$) is the creation operator for a particle on the $A$ ($B$) sublattice of the $m$th cell, and $c_{m,A}$ ($c_{m,B}$) is the corresponding annihilation operator. $v_m$ and $w_m$ characterize the intracell and intercell hopping amplitudes, respectively. This model describes a chiral chain with the Hamiltonian $H$ obeying $SH=-HS$. Here, the chiral operator $S=I_{L} \otimes \sigma_z$, where $\sigma_z$ is the Pauli matrix (sublattice space), and $I_{L}$ is a $L\times L$ identity matrix (unit-cell coordinate space). To preserve the chiral symmetry, we consider the quasi-periodic modulated intracell and intercell hoppings respectively with the amplitudes,
\begin{equation}
	v_m=v+{\Delta_m},\quad w_m=w+\gamma{\Delta_m},
	\label{eq2}
\end{equation}
with
\begin{equation}
	{\Delta_m}={\Delta}\cos(2\pi\beta m+\phi).
	\label{eq3}
\end{equation}
Here, $v$ and $w$ are the site-independent intracell and intercell tunneling energies, $\Delta$ is the strength of the incommensurate modulation, $\gamma$ is the ratio of quasi-periodic modulations of intracell to intercell tunneling (here we consider $\gamma>1$), $\beta$ is an irrational number to ensure the incommensurate modulation, and {$\phi\in[0,2\pi)$ is an arbitrary phase, which is used to create different quasi-periodic modulation configurations.} In the clean case ($\Delta=0$), the Hamiltonian Eq.~(\ref{eq1}) reduces to a standard SSH model \cite{Su1979}. {When the intracell hopping amplitude $v$ exceeds the intercell hopping amplitude $w$, the system undergoes a topological phase transition accompanied by the vanishing of the zero-energy edge modes and the nontrivial winding number.} The generalized SSH model described by Eq. (\ref{eq1}) can be realized by cold atoms in the momentum lattice. One can adjust the Bragg-coupling parameters between adjacent momentum-space sites to realize the off-diagonal quasi-periodic modulations \cite{Bo2021}. In the following, we will discuss the topological and localization properties of the SSH model with the quasi-periodic modulated hopping, respectively, and elaborate on the relationship between the reentrant localization transition and the TAI/RTAI. We set $w=1$ as the unit energy, $\beta=\frac{\sqrt{5}-1}{2}$ as the golden ratio, ${v\ge 0}$, $\Delta\ge 0$, and $\gamma=2$ for our numerical calculations under open boundary conditions (OBCs).
\begin{table*}
\begin{centering}
\begin{tabular}{|c|c|c|c|}
\hline
Site-independent intracell strength ($v$)                   & \multicolumn{1}{l|}{Disorder strength($\Delta$)} & Zero modes & Type of phases                                                                                              \\ \hline
$0<v<0.6$                  & $\Delta>0$                                       & Yes                                & Topological Insulator                                                                                       \\ \hline
\multirow{3}{*}{$0.6<v<1$} & $0<\Delta<\Delta_1$                                & Yes                                & \multirow{3}{*}{\begin{tabular}[c]{@{}c@{}}RTAI\end{tabular}} \\ \cline{2-3}
                                    & $\Delta_1<\Delta<\Delta_2$                       & No                                 &                                                                                                             \\ \cline{2-3}
                                    & $\Delta>\Delta_2$                                & Yes                                &                                                                                                             \\ \hline
\multirow{2}{*}{$v>1$}     & $0<\Delta<\Delta_3$                                & No                                 & \multirow{2}{*}{TAI}                                                        \\ \cline{2-3}
                                    & $\Delta>\Delta_3$                                & Yes                                &                                                                                                             \\ \hline
\end{tabular}
\end{centering}
\caption{Summary of the topological properties of the model (\ref{eq1}). Other parameters: $\gamma=2,w=1$. Here $\Delta_1$ and $\Delta_2$ are the topological phase transition points for $0.6<v<1$, as determined by Eqs.(\ref{eq8_1}) and (\ref{eq8_2}). And $\Delta_3$ is also the topological phase transition points for $1<v$, as determined by Eq.(\ref{eq8_2}).}
\label{tab1}
\end{table*}
\section{Topological phase transition}
\begin{figure}[h]
\includegraphics[width=0.45\textwidth]{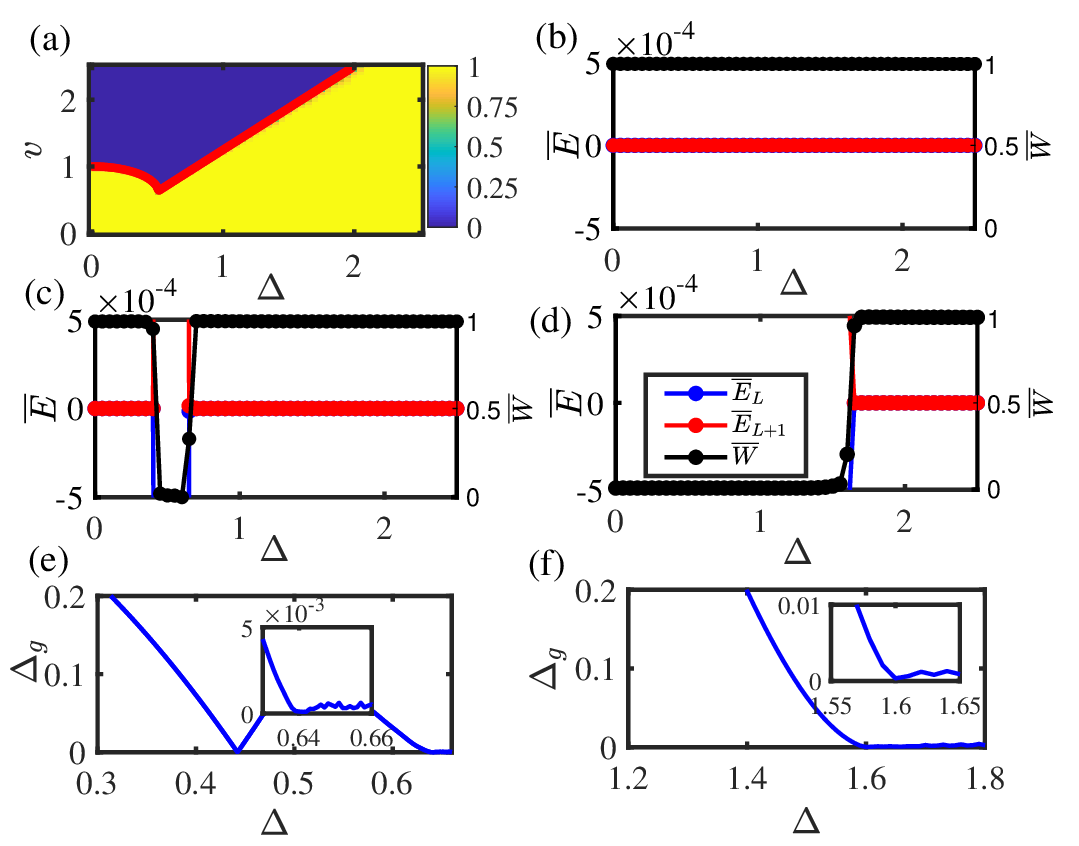}
\caption{(Color online) (a) Disorder-averaged winding number $\overline{W}$ as a function of the modulation strength $\Delta$ and the amplitude of the intracell hopping $v$ with $L=200$. The colorbar shows the value of disorder-averaged winding number $\overline{W}$. Two red dot lines represent the analytic critical lines Eqs. (\ref{eq8_1}) and (\ref{eq8_2})  for the divergence of the localization length $\lambda$ of the $L$-th eigenstate. Two disorder-averaged energies $\overline{E}_L$ and $\overline{E}_{L+1}$ in the center of the spectrum, and the disorder-averaged winding number $\overline{W}$ as a function of the modulation strength $\Delta$ under OBCs with  (b) $v=0.2$, (c) $v=0.8$, and (d) $v=2$, respectively. The bulk gap $\Delta_g$ as a function of $\Delta$ under PBCs with (e) $v=0.8$ and (f) $v=2$, respectively. Here, we take $N_c=100$ disorder realizations.}
\label{Fig2}
\end{figure}

\begin{figure}[h]
	\includegraphics[width=0.5\textwidth]{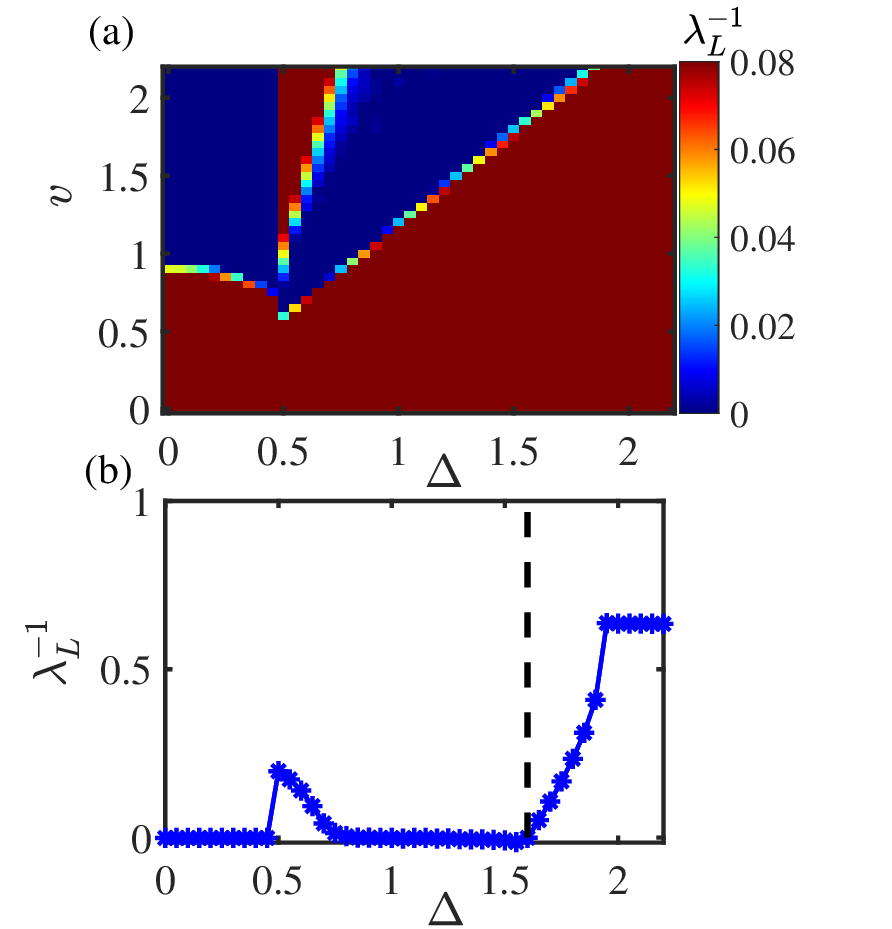}
	\caption{(Color online)(a)$\lambda_L^{-1}$ of the $L$-th eigenstate as a function of $v$ and $\Delta$ for $L=600$ and $N_c=100$ disorder realization. The colorbar shows the value of $\lambda_L^{-1}$.{(b) $\lambda_L^{-1}$ as a function of $\Delta$ for $v=2$ with $L=600$ and $N_c=100$ disorder realization. The black dash line corresponds to the topological transition point Eq. (\ref{eq8_2}).}}
	\label{Fig3}
\end{figure}

One can apply the open-bulk winding number to obtain the topological phase diagram of the SSH model without {the} translational symmetry. For a given modulation configuration, we can solve the Hamiltonian as $H|\psi^{n}\rangle=E_n|\psi^{n}\rangle$ with $E_n\ge 0$ and $|\tilde{\psi}^{n}\rangle =S|\psi^{n}\rangle$ corresponding to an eigenvector with $-E_n$, where the entries of the chiral operator are $S_{m\alpha,h\delta}=\delta_{mh}(\sigma_z)_{\alpha\delta}$ with $m,h$ referring to the unit cell and $\alpha,\delta$ to the sublattice. We introduce an open-boundary $Q$ matrix given by
\begin{equation}
	Q=\sum_{n}^{\prime}\left(|\psi^{n}\rangle \langle \psi^{n}| - |\tilde{\psi}^{n}\rangle \langle \tilde{\psi}^{n}|\right),
	\label{eq4}
\end{equation}
where $\sum\limits_{n}^{\prime}$ is the sum over the eigenstates in the bulk spectrum without the edge modes.
The open-bulk winding number in real space is defined as \cite{Song2019PRL,Kitaev06}
\begin{equation}
	W_c = \frac{1}{2L^{\prime}}\mathrm{Tr}^{\prime}(SQ[Q,X]).
	\label{eq5}
\end{equation}
Here, $X$ is the coordinate operator with $X_{m\alpha,h\delta}=m\delta_{mh}\delta_{\alpha\delta}$, which means that $X$ is a diagonal matrix given by $X={\rm{diag}}(1,1,2,2,...,L,L)$. {And the chiral operator $S={\rm{diag}}(1,-1,1,-1,...,1,-1)$. To mitigate the boundary effects, the chain is divided into three intervals with lengths $l$, $L^{\prime}$, and $l$, where $L=L^{\prime}+2l$. The focus of the analysis is on the middle interval with length $L^{\prime}$. The symbol $\mathrm{Tr}^{\prime}$ represents the trace operation over this middle interval of length $L^{\prime}$. $W_c$ serves well for disordered systems as it does not require the translation invariance and is quantized to an integer in the limit of $L \rightarrow\infty$. In practical calculations, a moderate system size is sufficient. In the context of Fig. \ref{Fig2}, a chain of length $L=200$ is chosen.} Furthermore, we define the disorder-averaged winding number $\overline{W}=1/N_c\sum_{c=1}^{N_c} W_c$, where $N_c$ is the number of the disorder realization by choosing different phases $\phi$. The topological properties of disordered system can also be characterized by the Bott index. And the detailed results can be found in Appendix D.

Figure \ref{Fig2}(a) shows the topological phase diagram on the $v$-$\Delta$ plane obtained by numerically computing $\overline{W}$. As shown in Fig. \ref{Fig2}(a), one can divide the topological phase diagram into three regions{\it, i.e.}, $v<0.6$, $0.6<v<1$, and $v>1$.
{When $v<0.6$, the SSH model is topologically nontrivial, and this feature is robust against the quasi-periodic modulations. It implies that in this regime, an arbitrary quasi-periodic modulation amplitude $\Delta$ will not break the nontrivial topology of the system. For the case $v=0.2$, with the increase of $\Delta$, $\overline{W}$ keeps unit, and the values of $\overline{E}_L$ and $\overline{E}_{L+1}$ remain zero, as shown in Fig. \ref{Fig2}(b). Here, $\overline{E}_{n}=1/N_c\sum_{c=1}^{N_c} E_n^c$ with $E_n^c$ being the $n$-th eigenenergy for a given modulation configuration.}
For $v \in (0.6,1)$, the system undergoes topological transitions among nontrivial-trivial-nontrivial regions as $\Delta$ increases. A clear picture can be obtained by the average winding number and the two disorder-averaged zero modes as a function of $\Delta$ for $v=0.8$, as shown in Fig. \ref{Fig2}(c). For small $\Delta$, $\overline{W}=1$ and $\overline{E}_L=\overline{E}_{L+1}=0$, which is topologically nontrivial. For $\Delta \in (0.45,0.65)$, the average winding number $\overline{W}$ jumps to $0$ and the values of the corresponding $\overline{E}_L$ and $\overline{E}_{L+1}$ break into nonzero pairs. It indicates that the modulation destroys the nontrivial topology. A modulation-induced topology emerge again in the $\Delta>0.65$ region, which is named as RTAI. {A detailed explanation for this process can be seen by the bulk gap $\Delta_g=\overline{E}_{L+1}-\overline{E}_{L}$ under period boundary conditions (PBCs), as shown in Fig.~\ref{Fig2}(e) with $v=0.8$. Initially, the system is in a topologically nontrivial phase characterized by a bulk gap. This topological gap is robust against weak disorder. However, as the disorder strength increases, a sufficiently strong disorder disrupts the topological order, leading to the closing of the bulk gap at a critical disorder strength of $0.45$. Within the interval $\Delta\in(0.45,0.65)$,  the system transitions to a topologically trivial phase with a corresponding trivial bulk gap.} For the region where $v>1$, the clean system exhibits topologically trivial characteristics. However, when the modulation amplitude surpasses a critical value, the system transitions into the TAI phase, which remains robust against the effects of the modulation. As illustrated in Fig. \ref{Fig2}(d) for the case with $v=2$ for $\Delta>1.6$, moderate modulations induce the emergence of disorder-averaged zero modes, coinciding with a sudden jump of the winding number $\overline{W}$ from $0$ to $1$. From Figs.~\ref{Fig2}(f) shown $\Delta_g$ as a function of $\Delta$ with $v=2$ under PBCs, it is evident that the energy spectrum undergoes a gapped-to-gapped transition at $\Delta \approx 1.6$, with the topological phase transition occurring at the point where the gap closes.
Both TAI and RTAI phases induced by the quasi-periodic modulations are robust against the disorder in our system.

According to Figs. \ref{Fig2}(b)-(d), in the topologically nontrivial regime, we find the emergence of the zero modes is always accompanied by a nonzero average winding number $\overline{W}=1$, and the zero modes are localized at the edges with a finite localization length. However, when the system enters the trivial regime, these edge modes vanish and bulk states emerge with the divergence of the localization length \cite{Hughes2014,Longhi2020}. Therefore, one can analytically obtain the topological phase {boundary} by studying the localization length of the zero modes. The Schr\"{o}dinger equation of the SSH model Eq. (\ref{eq1}) with zero modes, $H\psi=0$, is given by:
\begin{eqnarray}
w_m\psi_{m,B}+v_{m+1}\psi_{m+1,B}&=&0 \notag \\
v_m\psi_{m,A}+w_m\psi_{m+1,A}&=&0,
\label{eq6}
\end{eqnarray}
where $\psi_{m,A}$($\psi_{m,B}$) is the probability amplitude of the zero mode on the sublattice site $A(B)$ in the $m$-th lattice cell. By solving the coupled equations, one can obtain $\psi_{n+1,A}=(-1)^n\prod_{m=1}^{n} (v_m/w_m) \psi_{1,A}$, leading to the localization length $\lambda$ of the zero modes given by \cite{Longhi2020,J.K.2016}
\begin{eqnarray}
\lambda^{-1}&=&-\lim_{L\to\infty}\frac{1}{L}\ln\left|\frac{\psi_{L+1,A}}{\psi_{1,A}}\right| \notag \\
&=&\lim_{L\to\infty}\frac{1}{L}\left|\sum_{m=1}^{L}\ln{\frac{|1+\gamma\Delta\cos(2\pi \beta m)|}{|v+\Delta\cos(2\pi \beta m)|}}\right|.
\label{eq7}
\end{eqnarray}
The divergence of the localization length $\lambda$, \emph{i.e.}, $\lambda^{-1}\to 0$, gives the topological phase transition boundaries (see Appendix A for the derivation)
\begin{subequations}
\small
\begin{align}
{{1+\sqrt{1-\gamma^2\Delta^2}}}=&{v+\sqrt{v^2-\Delta^2}}&  \Delta<\frac{1}{\gamma}\ {\rm{and}} \ \Delta<v \label{eq8_1}\\
\Delta=&\frac{2\gamma}{1+\gamma^2}v  & {\frac{1}{\gamma}<\Delta<v.} \label{eq8_2}
\end{align}
\end{subequations}
The analytic results are shown in Fig. \ref{Fig2}(a) marked by the red solid lines, which match our numerical results. {Based on the above analysis and numerical results, the topological properties of the system can be summarized in Table \ref{tab1}.}

One can numerically calculate the value of $\lambda_L^{-1}$ for the $L$-th eigenstate, which is also known as the Lyapunov exponent, {defined as} \cite{Hughes2014,WZH2022PRB,MacKinnon1983}
\begin{eqnarray}
\lambda_L^{-1} &=&\lim_{L\to\infty}\frac{1}{L}\ln||T||,
	\label{eq12}
\end{eqnarray}
where $\|T \|$ denotes the norm of the total transfer matrix $T=\prod_{m=2}^{L}T_mT_1$ with
\begin{eqnarray}
T_m&=&\left[\begin{array}{cc}
\frac{E_L^2-v_m^2-w_{m-1}^2}{v_mw_m} & -\frac{w_{m-1}v_{m-1}}{v_mw_m} \\
1 & 0 \\
\end{array}
\right],
	\label{eq13}
\end{eqnarray}
and
\begin{eqnarray}
T_1 &=& \left[\begin{array}{cc}
\frac{E_L^2-v_m^2}{v_mw_m}& -1\\
1 & 0\\
\end{array}
\right].
	\label{eq14}
\end{eqnarray}
Figure.~\ref{Fig3}(a) shows the $\lambda_L^{-1}$ for the $L$-th eigenstate as a function of $v$ and $\Delta$ with $2L=1200$ and $N_c=100$. {In a topologically nontrivial regime, the $L$-th state represents a nontrivial zero mode with finite localization length. However, in a topologically trivial phase, the $L$-th state denotes a conventional bulk state. The diverging lines indeed match our topological phase boundaries in Fig. \ref{Fig2}(a). Note that the red triangle region in Fig. \ref{Fig3}(a) signifies a topologically trivial area. Through the discussion in the following text, this region corresponds to a fully localized phase where all states exhibit finite localization lengths. Therefore, within Fig. \ref{Fig3}(a), the presence of a red triangle region with $\lambda_L^{-1}>0$ suggests that the $L$-th eigenstate in the topologically trivial domain is a localized bulk mode.} Here, when the value of $\lambda_L^{-1}$ exceeds $0.08$, we represent it using a single color, which serves the practical purpose of highlighting the topological phase transition points in Fig. \ref{Fig3}(a). {As an example, we choose $v=2$ to plot $\lambda_L^{-1}$ as a function of $\Delta$ in Fig.~\ref{Fig3}(b). For $\Delta$ within the range $(0.5,0.8)$, $\lambda_L^{-1}>0$, indicating that the $L$-th eigenstate is a trivial bulk state exhibiting localized features. When $\Delta>\Delta_c=1.6$, the $L$-th eigenstate corresponds to a topologically nontrivial edge mode. The value $\Delta_c$ marks the transition point for the TAI phase, indicated by the black dashed line in Fig.\ref{Fig3}(b).}

\section{Localization phase diagram and Reentrant localization transition}

\begin{figure*}[tbp]
\centering
\includegraphics[width=1\textwidth]{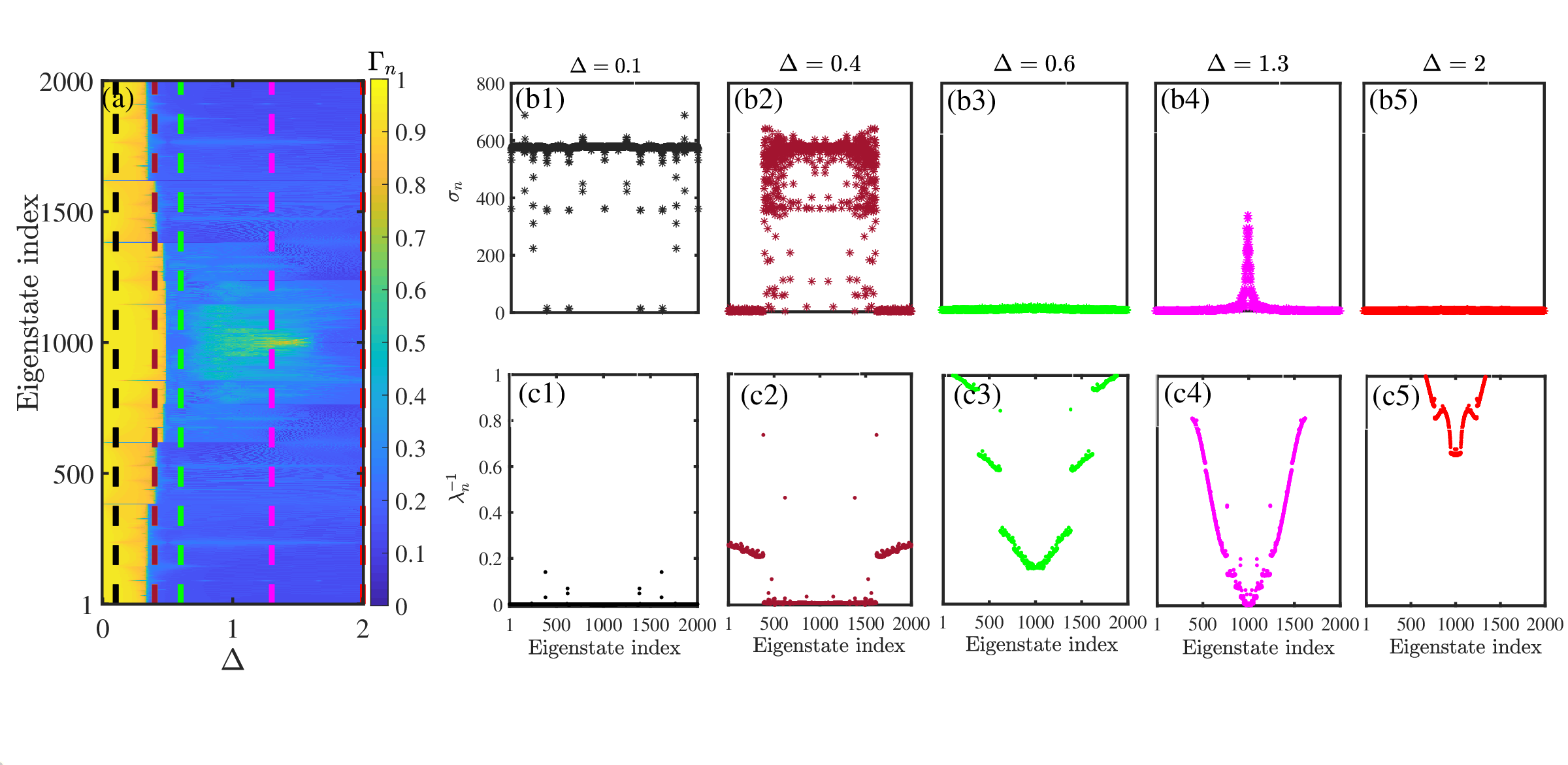}
\caption{(Color online)(a)The fractal dimension  $\mathrm{\Gamma}_n$ associated with the eigenstate indices as a function of $\Delta$. The colorbar indicates the magnitude of $\mathrm{\Gamma}_n$.
The vertical dashed lines of different colors correspond to $\Delta=0.1, 0.4, 0.6, 1.3$, and $2$, respectively. And the corresponding distributions of the $\sigma_n$ and $\lambda_n^{-1}$ as a function of the eigenstate index are shown in (b1)-(b5) and (c1)-(c5), respectively. Other parameters: $\phi=0$, $v=2$, and $L=1000$.}
\label{Fig4}
\end{figure*}

To obtain the localization properties of the system, we rely on the inverse participation ratio $(\rm{IPR})$ and the normalized participation ratio $(\rm{NPR})$, which are defined respectively as
\begin{eqnarray}
\mathrm{IPR}_n &=& \sum_{m=1}^{L}\sum_{\alpha}|\psi_{m,\alpha}^{n}|^4, \notag \\
\mathrm{NPR}_n &=& \left(2L\sum_{m=1}^{L}\sum_{\alpha}|\psi_{m,\alpha}^{n}|^4\right)^{-1},
\label{eq9}
\end{eqnarray}
where $\psi_{m,\alpha}^{n}$ is the probability amplitude of the $n$-th eigenstate on the sublattice site $\alpha$($\alpha=A\ $or$\ B$) in the $m$-th unit cell. It is known that ${\mathrm{IPR}_n}$ tends to zero(nonzero) and ${\mathrm{NPR}_n}$ is nonzero(zero) for the delocalized(localized) phases in the thermodynamic limit. To observe the overall localization transition of the system, we average the ${\mathrm{IPR}_n}$ and ${\mathrm{NPR}_n}$ over all eigenstates to obtain $\langle\rm{IPR}\rangle$ and $\langle\rm{NPR}\rangle$.

In order to see the localization transition point more clearly, we can use the fractal dimension, which is defined as:
\begin{eqnarray}
	\mathrm{\Gamma}_n &=& -\lim_{L\to\infty}\frac{\ln{\mathrm{IPR}_n}}{\ln{(2L)}}.
	\label{eq10}
\end{eqnarray}
In the large $L$ limit, $\Gamma_n \to 0$ for localized states. For the delocalized states, when $\Gamma_n \to 1$ for extended states and $0<\Gamma_n<1$ for the critical ones \cite{Wang2020PRL2}.

In order to further distinguish the localization states from the delocalized ones, we can also apply the standard deviations of the coordinates $\sigma_n$ and the localization length $\lambda_n$ of eigenstates with the eigenvalue $E_n$, respectively. The standard deviations of the eigenstate coordinates $\sigma_n$ are given by \cite{Yicaizhang2022}
\begin{eqnarray}
\sigma_n&=&\sqrt{\sum\limits_{i=1}^{2L}\left(i-\overline{i}\right)^2\left|\psi^n(i)\right|^2,}
	\label{eq11}
\end{eqnarray}
where $i$ is the lattice coordinate and $\overline{i}=\sum\limits_{i=1}^{2L}i\left|\psi^n(i)\right|^2$ is the center of mass position. {The} $\sigma_n$ contains the information about the spatial distribution of the eigenstates. While $\sigma_n$ is small for a localized state and larger for the extended state, {the $\sigma_n$} of the critical states is in between of them and exhibit fluctuation. The inverse of the localization length $\lambda^{-1}_n$ for the $n$-th eigenstate measures the average growth rate of the wave function and can be calculated in a similar way as Eq.(\ref{eq12}) for $L$-th eigenstate. The case $\lambda^{-1}_n>0$ corresponds to a localized state, and a delocalized state is characterized by $\lambda^{-1}_n=0$ \cite{yanxialiu2}. Figure. \ref{Fig4}(a) shows the fractal dimensions $\Gamma_n$ associated to eigenstate indices as a function of $\Delta$. It can be seen that some eigenstates are localized and the rest are delocalized in $0.3<\Delta<0.5$ and $0.7<\Delta<1.6$ regions. We discuss the localization features in different regimes by taking $\Delta=0.1, 0.4, 0.6, 1.3$, and $2$ as examples, which are marked by dashed lines with different colors shown in Fig.~\ref{Fig4}(a). The corresponding distributions of  $\sigma_n$ and $\lambda^{-1}_n$ as a function of the eigenstate indices $n$ are shown in Fig. \ref{Fig4}(b1)-(b5) and \ref{Fig4}(c1)-(c5), respectively. For $\Delta=0.1$, all the eigenstates are extended. As shown in Fig.~\ref{Fig4}(b1) and (c1), the standard deviations of almost all eigenstates $\sigma_n$ are stabilized at a large value and the corresponding values of the $\lambda^{-1}_n$ approach zero. When  $\Delta=0.4$, $\lambda^{-1}_n \to 0$ and the standard deviations of the eigenstates $\sigma_n$ display extended behaviors in the band-center region, as shown in the Fig.~\ref{Fig4}(b2) and (c2). And in the band-edge region, $\lambda^{-1}_n >0$ and $\sigma_n$ is very small corresponding to the localized properties. The coexistence of both delocalized and localized states is a typical characteristics of the intermediate phase hosing MEs. Further increasing $\Delta$ to $0.6$, all eigenstates are localized, where the standard deviations are stabilized at very small values and $\lambda^{-1}_n$ take finite values, as shown in the Fig.~\ref{Fig4}(b3) and (c3). In Fig.~\ref{Fig4}(b4) and (c4), we find that the values of $\sigma_n$ of some of eigenstates in the band-center region exhibit a relatively large fluctuation and the corresponding values of $\lambda^{-1}_n$ tend to zero, implying that some eigenstates in the band-center region reenter into the delocalization regime for $\Delta=1.3$.
When the modulation strength is large enough, such as $\Delta=2$ in Fig. \ref{Fig4}(b5) and (c5), the system is recovered into a fully localized regime. As $\Delta$ increases, the band of the spectrum shows a sequential transition between extended-intermediate-localized-intermediate-localized regions. These numerical results clearly show that the existence of two intermediate regions.

\begin{figure}[h]
\includegraphics[width=0.45\textwidth]{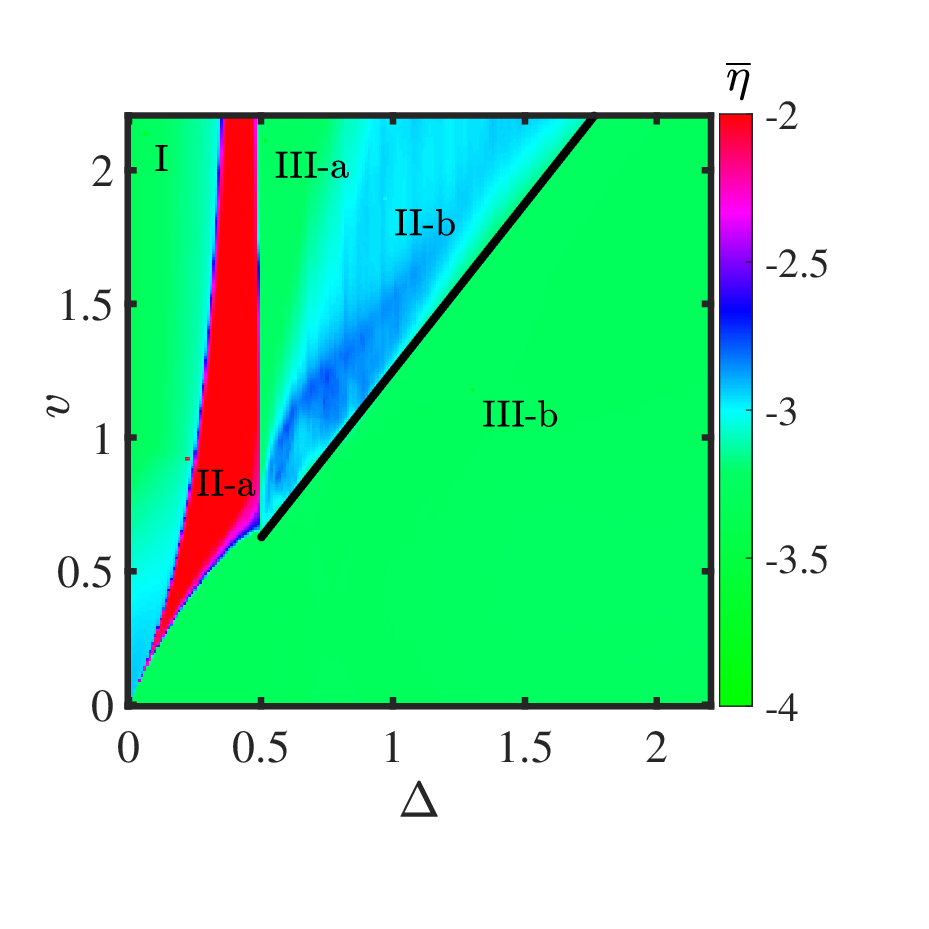}
\caption{(Color online) The localization phase diagram in the $v-\Delta$ plane obtained using the values of $\overline{\eta}$ for $L=1000$. The regions for delocalized, intermediate and localized phases are denoted by {\uppercase\expandafter{\romannumeral1}
}, \uppercase\expandafter{\romannumeral2}-a(\uppercase\expandafter{\romannumeral2}-b)
, and \uppercase\expandafter{\romannumeral3}-a(\uppercase\expandafter{\romannumeral3}-b)
, respectively.  The black lines represent the analytic critical lines Eq. (\ref{eq8_2}) for topological transition point. The colorbar represents the values of $\overline{\eta}$. All data in this figure are averaged over $100$ disorder realizations.}
\label{Fig5}
\end{figure}

To obtain the full localization phase diagram on the $\Delta-v$ plane, we define a disorder-average quantity $\overline{\eta}=1/N_c\sum_{c=1}^{N_c}\eta_c$ \cite{xiaoli}, which can clearly distinguish the intermediate region from the fully extended and localized regions in the phase diagram. Here,
\begin{equation}
\eta_c = \log_{10}\left[\langle\rm{IPR}\rangle\langle\rm{NPR}\rangle\right].
	\label{eq15}
\end{equation}
As shown in Fig. \ref{Fig5}, there are three phases in this system: I, II-a(II-b), and III-a(III-b) corresponding to the delocalized, intermediate and localized phases, respectively. For $v<0.6$, only one intermediate region exists in a finite range of $\Delta$. While for $v>0.8$, the phase diagram clearly shows two intermediate regions, marked in red and blue, separated by Phase III-a. Note that the blue region in the small $v$ and $\Delta$ limit corresponds to an extended but nonergodic phase, where it contains many critical states, that is also marked as a delocalized phase. A specific example with $v=0.2$ and $\Delta=0.08$ is considered in Appendix B. We also find that at $\Delta=0.5$ for a larger $v$, there is a localization transition point, which corresponds to a transition from the intermediate phase to the fully localized phase. Our numerical results have revealed that for larger values of $v$, such localization transition occurs at $\Delta = 1/\gamma$. This particular point aligns with the localization transition of the $L$-th eigenstate of the system shown in Fig. \ref{Fig3}(a).

Comparing the localization phase diagram Fig. \ref{Fig5} with the topological phase diagram Fig.~\ref{Fig2}(a), a noteworthy observation is that the reentrant phase transition closely tracks the RTAI/TAI phase transition described by Eq.(\ref{eq8_2}), as highlighted by the black solid line in Fig.~\ref{Fig5}. The intricate interplay of topological properties is crucial in governing the emergence of the reentrant localization transition. Phase III-b is identified as a topologically non-trivial phase (TAI/RTAI phase), characterized by the presence of topologically localized zero modes in the band-center region. In contrast, Phase III-a represents a topologically trivial phase lacking edge modes, and localized bulk modes emerge in the band-center region. Previous studies \cite{Hughes2014, DZhang2020, Meier2018} support the notion that localized bulk modes at the band center undergo a process of delocalization, evolving into topologically localized edge modes. This suggests that some initially localized bulk modes centered in the band-center region become delocalized with the increase of $\Delta$, giving rise to the emergence of an intermediate phase (Phase II-b) between Phase III-a and Phase III-b. Consequently, the emergence of the reentrant localization transition is closely tied to the TAI transition in our case. To further illuminate the correlation between reentrant localization and the TAI/RTAI phase transition, specific examples with $v=1.6$ and $1.9$ are considered (see details in Appendix C). {Based on the numerical results of Fig.~\ref{Fig5}, the
localization properties of the system can be summarized in Table \ref{tab2}.}
\begin{table*}
\begin{centering}
\begin{tabular}{|l|l|}
\hline
Site-independent intracell strength ($v$)             & Localization transition processes \\ \hline
$0< v < 0.6$  &           \multicolumn{1}{c|}{I$\rightarrow$II-a$\rightarrow$III-b}              \\ \hline
$0.6< v < 0.8$ &           \multicolumn{1}{c|}{I$\rightarrow$II-a$\rightarrow$II-b$\rightarrow$III-b} \\ \hline
$0.8< v$      &            \multicolumn{1}{c|}{I$\rightarrow$II-a$\rightarrow$III-a$\rightarrow$II-b$\rightarrow$III-b}     \\ \hline
\end{tabular}
\end{centering}
\caption{{Summary of the localization properties of the model (\ref{eq1}). Other parameters:$\gamma = 2,w = 1$. Here I, II-a(II-b), and III-a(III-b) corresponding to the delocalized, intermediate and localized phases, respectively.}}
\label{tab2}
\end{table*}
\section{Dynamical Detection}

\begin{figure}[tbp]
	\includegraphics[width=0.5\textwidth]{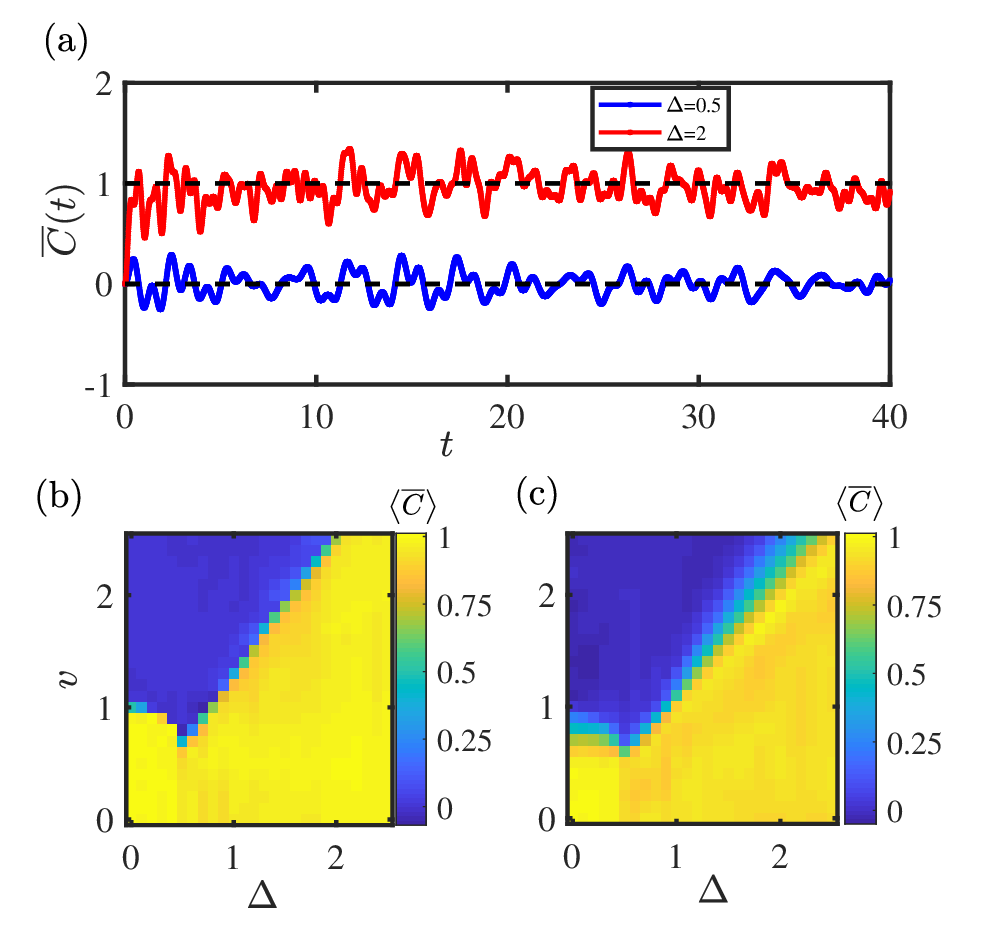}
	\caption{(Color online) (a) Dynamics of $\overline{C}(t)$ with $L=100$, $v=2$ and $N_c=100$ disorder realizations for $\Delta=0.5$ and $\Delta=2$, respectively. $\langle \overline{C} \rangle$ as a function of the modulation strength $\Delta$ and the amplitude of the intracell hopping $v$ for (b) $L=100$ and (c) $L=10$ over $100$ disorder realizations. The colorbar indicates the magnitude of $\langle \overline{C} \rangle$ in (b) and (c).}
	\label{Fig6}
\end{figure}

\begin{figure}[tbp]
	\includegraphics[width=0.5\textwidth]{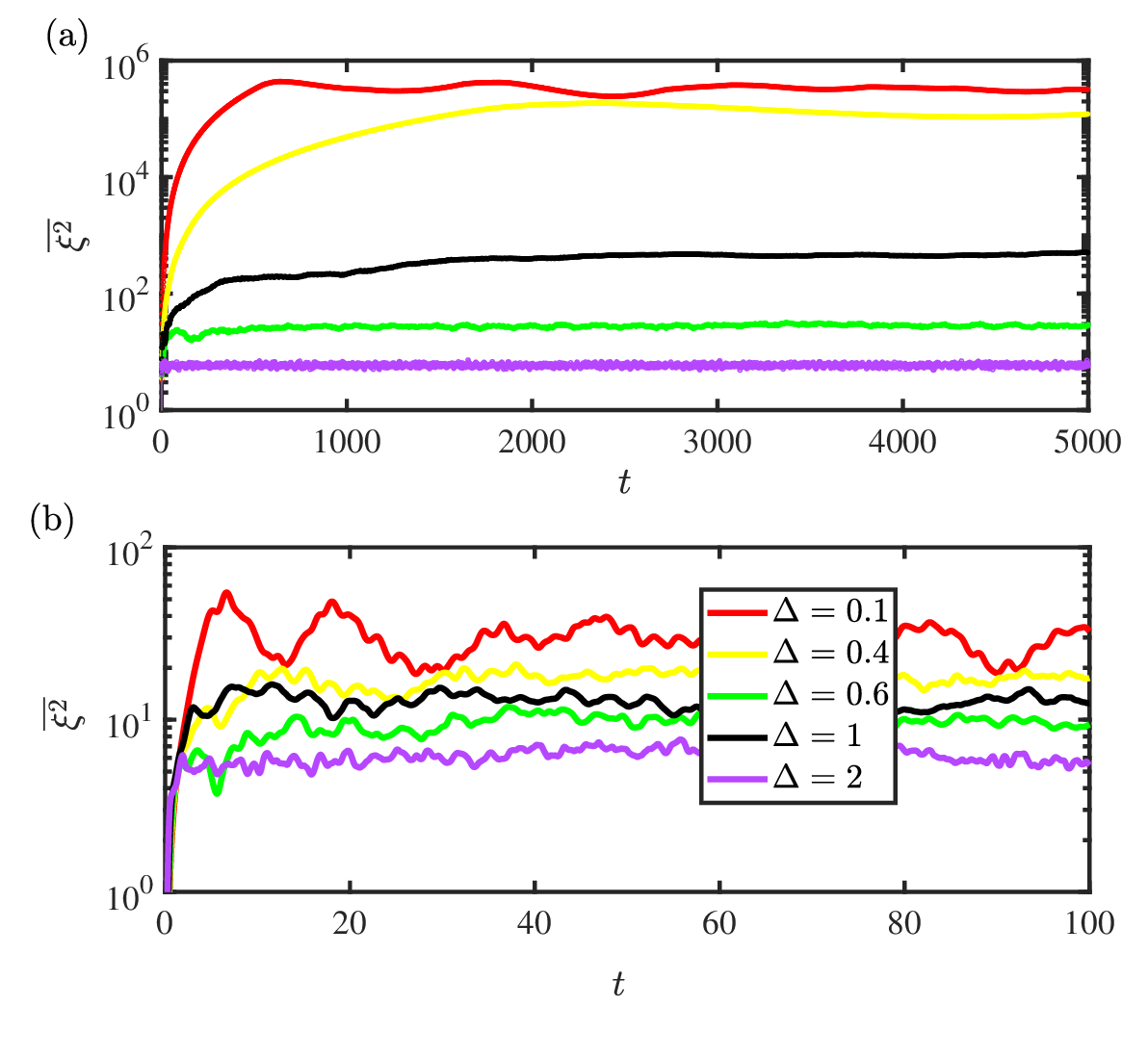}
	\caption{(Color online) The mean-square displacement $\overline{\xi^2}(t)$ as a function of the time $t$ for $\Delta=0.1, 0.4, 0.6, 1$, and $2$ with (a) $L=1000$ and (b) $L=10$ under $N_c=100$ disorder realizations. Here, $v=2$.}
	\label{Fig7}
\end{figure}
\begin{figure}[tbp]
	\includegraphics[width=0.5\textwidth]{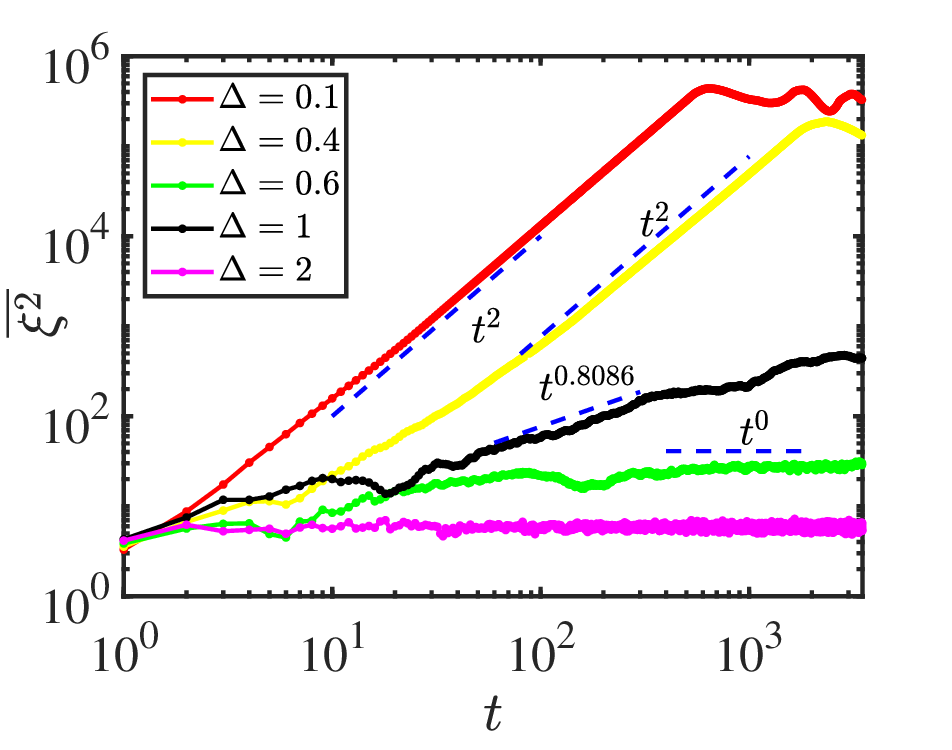}
	\caption{(Color online) The log-log plot of the mean-square displacement $\overline{\xi^2}(t)$ as a function of the time $t$ for $\Delta=0.1, 0.4, 0.6, 1$, and $2$ with $v=2$ and $L=1000$ under $N_c=100$ disorder realizations.}
	\label{Fig8}
\end{figure}

To realize our model, one can implement the 1D momentum lattice technique in cold atom experiments. Discrete momentum states can be coupled by pairs of Bragg lasers. By engineering the frequencies $\omega_h$ of the multicomponent Bragg lasers, the counter-propagating laser pairs drive a series of two-photon Bragg resonance transitions, which couple the adjacent momentum states separated by $2\hbar k$, where $k$ is the wave number. The off-diagonal modulation, in our case, can be individually tuned by adjusting the amplitude of the Bragg beam with frequency $\omega_h$ for the function $v_m$ and $w_m$. Here, $\omega_h$ is tuned to the two-photon resonance between the corresponding momentum states via an acousto-optic modulator. In current experiments, by using $^{87}$Rb or $^{133}$Cs atoms, the typical system size is $\sim 20$ sites. In the following, we choose $L=10$ for our numerical simulation. We also calculate the dynamical detection with a large system size for comparison.

To dynamically investigate the topological properties of the quasi-periodic modulated SSH model, one can measure the mean chiral displacement \cite{Meier2018,Scherg2018}. We define the single-shot expectation value of the mean chiral displacement operator in a given modulation configuration as \cite{Meier2018,SNL2022}
\begin{eqnarray}
	C_c(t) = 2\langle\psi(t)|SX'|\psi(t)\rangle,
	\label{eq16}
\end{eqnarray}
where $|\psi(t)\rangle=e^{-iHt}|\psi(0)\rangle$ is the time-evolved wave function. The wave function $|\psi(0)\rangle$ is initially localized at the center site of the chain. In cold atom experiments, the momentum lattice technique allows for the initial confinement of the entire atomic population to a single momentum site in the lattice center. Here, $X'=X-(L/2+1)$ \cite{Meier2018}. The dynamics of the disorder-average mean chiral displacement $\overline{C}(t) = 1/N_c\sum_{c=1}^{N_c} C_c(t)$ generally exhibit a transient, oscillatory behavior. To eliminate the oscillation, we take their time-average $\langle \overline{C} \rangle$, which converges to the corresponding winding number $\overline{W}$ \cite{Meier2018,SNL2022}. Figure \ref{Fig6}(a) displays the dynamics of $\overline{C}(t)$ for both weak ($\Delta=0.5$) and strong ($\Delta=2$) modulations for $v=2$ with $L=100$. We can see that the dynamics of $\overline{C}(t)$ show transient and oscillatory processes and eventually converge to their corresponding $\overline{W}$ (marked by the black dashed lines, respectively) in long time limits. Moreover, we can obtain the topological phase diagram through the $\langle \overline{C} \rangle$ in the $v-\Delta$ plane shown in Figs.~\ref{Fig6}(b) with $L=100$ and (c) with $L=10$. By comparing the dynamical evolution behaviors of different system sizes, one can find that the mean chiral displacement $\langle \overline{C} \rangle$ can effectively characterize the topological properties even in a small-sized system.

\begin{figure}[tbp]
	\includegraphics[width=0.5\textwidth]{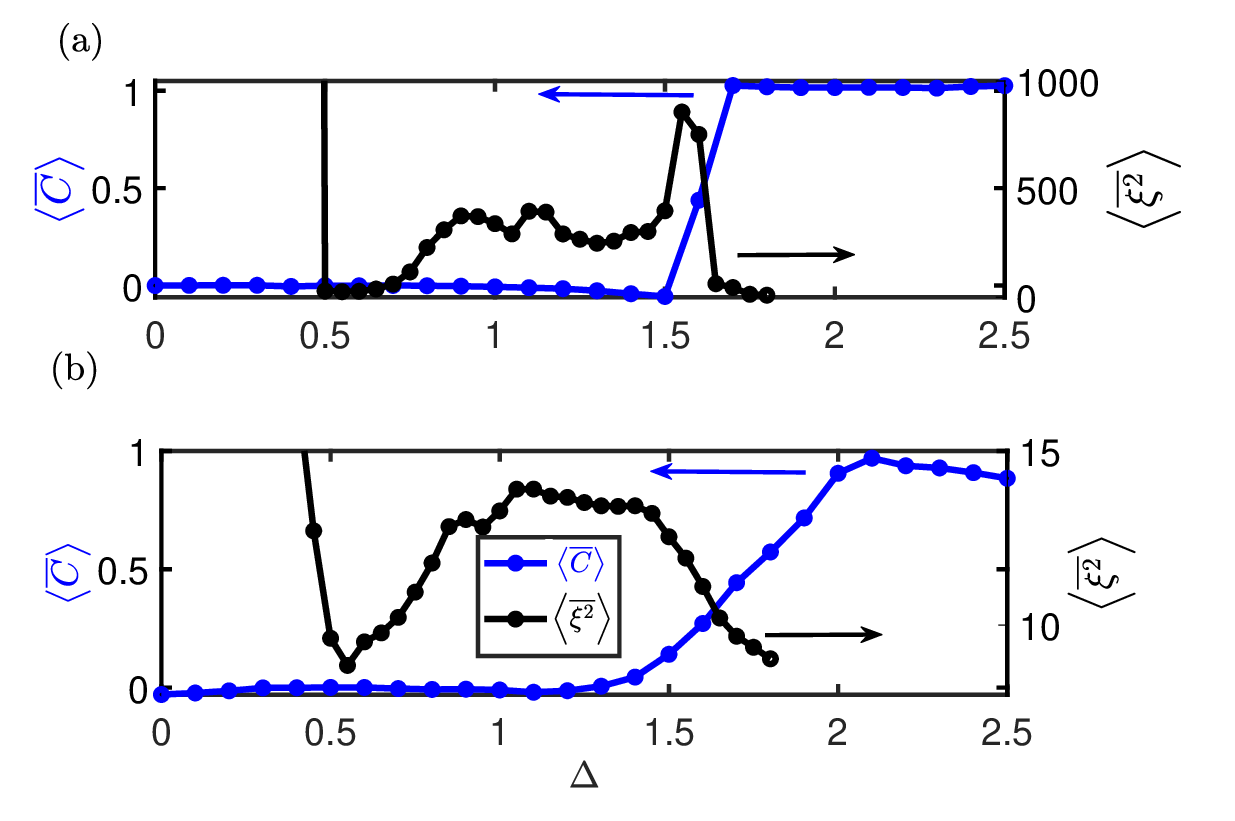}
	\caption{(Color online) The mean square displacement $\left<\overline{\xi^2}\right>$ (black line) and the mean chiral displacement $\left<\overline{C}\right>$ (blue line) as a function of the modulation strength $\Delta$ with (a) $L=1000$ and (b) $L=10$, respectively. Here, $v=2$ and all data in this figure are averaged over $100$ disorder realizations.}
	\label{Fig9}
\end{figure}

To characterize the signatures of the multiple localization transitions, we can calculate the mean-square displacement for a given modulation realization defined as \cite{Padhan2022PRB,Scherg2018,ZHXU2020}
\begin{eqnarray}
	\xi^2_c&=&\sum_{i}\left(i-i_0\right)^2\left|\psi_i(t)\right|^2,
	\label{eq18}
\end{eqnarray}
where the entire atomic population is initially localized at the center of the lattice $i_0$. We can use the behavior of the disorder-averaged mean-square displacement $\overline{\xi^2}$ in the long-time evolution to verify the localization properties of the system \cite{Padhan2022PRB}. Here $\overline{\xi^2}=1/N_c\sum_{c=1}^{N_c}\xi^2_c$. As shown in Fig.~\ref{Fig7}(a), $\overline{\xi^2}$ saturates to different values after a long time evolution with various $\Delta$ for $v=2$ and $L=1000$, which can display the multiple localization transitions. According to the localization phase diagram in Fig.~\ref{Fig5}, the system is in the extended, intermediate, localized, intermediate, and localized phases for $\Delta=0.1, 0.4, 0.6, 1,$ and $2$, respectively. The corresponding saturation values of $\overline{\xi^2}$ exhibit different features. For $\Delta=0.1$, the system is in the extended phase with the highest saturation values of $\overline{\xi^2}$. For $\Delta=0.4$ and $1$, the system is in the intermediate phase, and the saturation values of $\overline{\xi^2}$ are larger than those at $\Delta=0.6$ and $2$, where the system is in the localized phase. To examine the impact of size on the system, we also calculate $\overline{\xi^2}$ with different $\Delta$ for $L=10$, as shown in Fig.~\ref{Fig7}(b). It is evident that, even in the small size, the saturation value of $\overline{\xi^2}$ after a long evolution can still reflect the multiple localization transitions of the system.

In order to clearly characterize the signatures of localization transitions, we added a log - log plot of the mean-square displacement $\overline{\xi^2}$ as a function of the time $t$ for different $\Delta$ in Fig.~\ref{Fig8}. The values of $\overline{\xi^2}$ grow in a power-law form of time given by $\overline{\xi^2} \propto t^{\mu}$ during the expansion process. From Fig.~\ref{Fig8}, one can see that when $\Delta$=$0.1$, the value of $\overline{\xi^2}$ displays a ballistic diffusion with $\mu=2$, which indicates the system is in the extended phase. For $\Delta$=$0.4$, the dynamical behavior in the intermediate region, with the mixture of extended and localized states, displays $\mu=2$, but with a lower saturation value than the extended case \cite{ZHXU2020}. When $\Delta$=$0.6$ and $2$, $\mu=0$, which indicates the system is in the localized phase. When $\Delta$=$1$, the value of $\overline{\xi^2}$ displays subdiffusive behavior with $\mu=0.8086$, and a much lower saturation value for $\overline{\xi^2}$ than in the extended phase. This indicates that the system is in the intermediate region, with a mixture of critical and localized states. According to the behavior of the disorder-averaged mean-square displacement, we can find that the system undergoes a typical reentrant localization transition process.

To detect the topological transition associated with the localization transition in the expansion dynamics, we present a plot of the time-averaged mean-square displacement $\langle \overline{\xi^2}\rangle$  and $\langle \overline{C} \rangle$ as a function of $\Delta$ for $v=2$ with $L=1000$ and $10$, shown in Fig.~\ref{Fig9}(a) and (b), respectively.
In conventional disordered systems, a consistent trend is the decrease in mean-square displacement as disorder strength increases, indicating increased localization. However, in contrast, Fig.~\ref{Fig9} illustrates an unconventional behavior: $\langle \overline{\xi^2}\rangle$ undergoes an anomalous shift as disorder strength increases. This indicates an anomalous change in the system's localization degree and suggests the possibility of a reentrant localization transition within the system. Our numerical and analytical discussions verify the occurrence of a reentrant localization phenomenon in the system. As seen in Fig.~\ref{Fig9}, when $\Delta>1.5$, $\langle \overline{\xi^2}\rangle$ decreases rapidly, indicating that all the states are localized once again and the reentrant localization emerges. At the same time, the system entered the TAI region, as evidenced by a jump of $\langle \overline{C} \rangle$ from $0$ to $1$. These results suggest that reentrant localization occurs in association with the TAI transition. A comparison with the results in Fig.~\ref{Fig9}(b) reveals a similar transition process for a small-sized system $L=10$. The above analyses indicate that the expansion dynamics can be used to detect the consistency of the TAI and reentrant localization transitions.

\section{Summary}
In this study, we investigated the topological and localization properties of a generalized SSH model with off-diagonal quasi-periodic modulations. Contrary to the conventional case where sufficiently strong disorder always destroys the topologically nontrivial properties, a suitable choice of disorder structure can induce the emergence of the robust topological phase in the case of sufficiently strong disorder. In particular, we found that the off-diagonal quasi-periodic modulations can induce the emergence of the stable TAI and RTAI phases. Furthermore, we investigated the localization properties of our system. Comparing the topological phase diagram and the localization transition, we find that a reentrant localization transition always accompanies the TAI/RTAI transition. It implies that topology properties are crucial in establishing the reentrant localization transition. Finally, we employed wave-packet dynamics in different parameter regimes to characterize and detect the TAI and reentrant localization.
Our findings can be simulated in cold atom systems by applying the momentum lattice technigue.

\section*{Appendix A: Derivation of the topological phase transition point}
\setcounter{equation}{0}
\setcounter{figure}{0} \setcounter{table}{0} %
\renewcommand{\theequation}{A\arabic{equation}}
\renewcommand{\thefigure}{A\arabic{figure}}
\renewcommand{\thetable}{A\Roman{table}}
In this Appendix, we present a detailed derivation of the expression of Eqs. (\ref{eq8_1}) and (\ref{eq8_2}) in the main text.
First, the Eq. (\ref{eq7}) can be simplified as follows
\begin{eqnarray}
\small
\lambda^{-1}&=&\lim_{L\to\infty}\frac{1}{L}\Bigg|\sum_{m=1}^{L}\big[\ln{|1+\gamma\Delta\cos(2\pi \beta m)|}\notag\\
&-&\ln{|v+\Delta\cos(2\pi \beta m)|}\big]\Bigg|.
\label{A1}
\end{eqnarray}
According to Weyl's equidistribution theorem \cite{Weyl1916,Choe1993}, we can use the ensemble average to evaluate the above expression
\begin{eqnarray}
\small
\lambda^{-1}&=&\left|\frac{1}{2\pi}\int^{\pi}_{-\pi}(\ln{|1+\gamma{\Delta}\cos q|}-\ln{|v+{\Delta}\cos q|})dq \right|\notag\\
&=&\Big|\frac{1}{2\pi}\int^{\pi}_{-\pi}\ln{|1+\gamma{\Delta}\cos q|}dq\notag\\
&-&\frac{1}{2\pi}\int^{\pi}_{-\pi}\ln{|v+{\Delta}\cos q|}dq\Big|.
\label{A2}
\end{eqnarray}
The first part of the integration of the Eq.(\ref{A2}) can be performed straightforwardly as
\begin{equation}
\frac{1}{2\pi}\int^{\pi}_{-\pi}\ln{|1+\gamma{\Delta}\cos q|}dq=\begin{cases}\ln{\frac{1+\sqrt{1-\gamma^2\Delta^2}}{2}} \quad & 1>\gamma\Delta \\
	\ln{\frac{\gamma\Delta}{2}} \quad & 1<\gamma\Delta\\
\end{cases}
\label{A3}
\end{equation}
and the second part is
\begin{equation}
\small
\frac{1}{2\pi}\int^{\pi}_{-\pi}\ln{|v+{\Delta}\cos q|}dq=\begin{cases}\ln{\frac{v+\sqrt{v^2-\Delta^2}}{2}} \quad & v>\Delta, \\
	\ln{\frac{\Delta}{2}} \quad & v<\Delta. \\
	\end{cases}
\label{A4}
\end{equation}
Combining the results (\ref{A3}) and (\ref{A4}), there are four possible situations:
\begin{figure}[tbp]
 \includegraphics[width=0.5\textwidth]{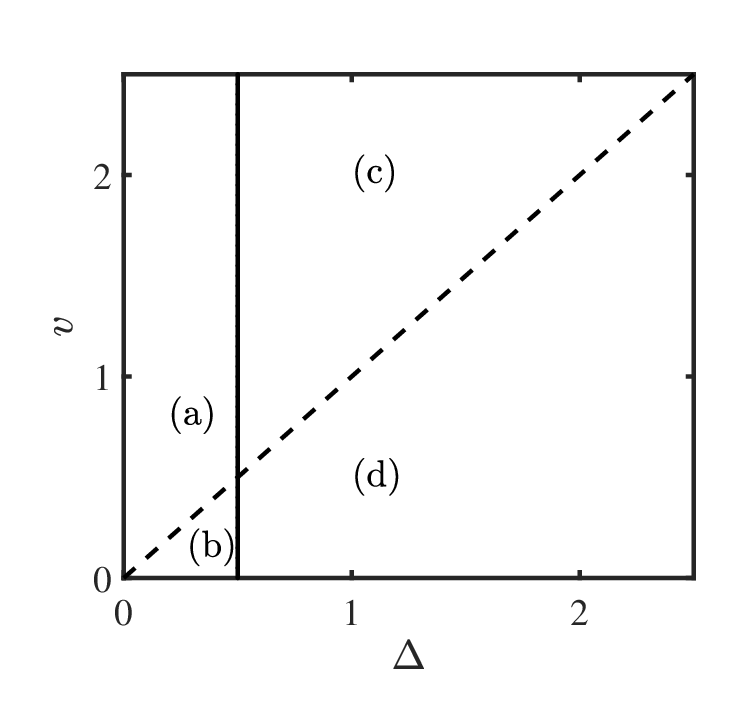}
  \caption{(Color online) The four regions: (a), (b), (c) and (d) in the $v-\Delta$ plane.}
\label{Fig10}
\end{figure}
(a)For $\Delta<\frac{1}{\gamma}$ and $\Delta<v$, we have
\begin{eqnarray}
\lambda^{-1}=\left|\ln{\frac{{1+\sqrt{1-\gamma^2\Delta^2}}}{v+\sqrt{v^2-\Delta^2}}}\right|.
\label{A5}
\end{eqnarray}
(b)For $\Delta<\frac{1}{\gamma}$ and $\Delta>v$, we have
\begin{eqnarray}
\lambda^{-1}=\left|\ln{\frac{{1+\sqrt{1-\gamma^2\Delta^2}}}{\Delta}}\right|.
\label{A6}
\end{eqnarray}
(c)For $\Delta>\frac{1}{\gamma}$ and $\Delta<v$, we have
\begin{eqnarray}
\lambda^{-1}=\left|\ln{\frac{\gamma\Delta}{v+\sqrt{v^2-\Delta^2}}}\right|.
\label{A7}
\end{eqnarray}
(d)For $\Delta>\frac{1}{\gamma}$ and $\Delta>v$, we have
\begin{eqnarray}
\small
\lambda^{-1}=\left|\ln\gamma\right|.
\label{A8}
\end{eqnarray}
According to the above results, the $v-\Delta$ plane can be divided into four regions by two lines $v=\Delta$ and  $\Delta=\frac{1}{\gamma}$ (Here, we consider $\gamma>1$), as shown in Fig.~\ref{Fig10}. $\lambda^{-1}=0$ is the topological phase transition point. Since $\gamma>1$, we can get $\left|\ln\gamma\right|>0$. No topological phase transition exist in region (d). For case (b) ($v<\Delta<\frac{1}{\gamma}$), we have
\begin{eqnarray}
{{1+\sqrt{1-\gamma^2\Delta^2}}}&=&\Delta,
\label{A9}
\end{eqnarray}
which contradicts the preliminary condition ($\Delta<\frac{1}{\gamma}$). Hence, the case (b) is also excluded. Combining the results (\ref{A5}) and (\ref{A7}), we obtain
\begin{small}
\begin{equation}
{{1+\sqrt{1-\gamma^2\Delta^2}}}={v+\sqrt{v^2-\Delta^2}}\ \Delta<\frac{1}{\gamma}\ {\rm{and}}\ \Delta<v,
\label{A11}
\end{equation}
\end{small}
and
\begin{eqnarray}
{\gamma\Delta}&={v+\sqrt{v^2-\Delta^2}}\ \frac{1}{\gamma}<\Delta<v.
\label{A12}
\end{eqnarray}
Here, $\Delta$ and $v$ are positive numbers. The Eq. (\ref{A12}) can be further simplified as
\begin{eqnarray}
\Delta&=\frac{2\gamma }{1+\gamma^2}v.
\label{A13}
\end{eqnarray}
Combined with the above results, we obtain the topological transition boundary Eqs. (\ref{eq8_1}) and (\ref{eq8_2}) in the main text.

\section*{Appendix B: Localization properties in small $v$ and $\Delta$ limits}
\setcounter{equation}{0}
\setcounter{figure}{0} \setcounter{table}{0} %
\renewcommand{\theequation}{B\arabic{equation}}
\renewcommand{\thefigure}{B\arabic{figure}}
\renewcommand{\thetable}{B\Roman{table}}
\begin{figure}[h]
\includegraphics[width=0.4\textwidth]{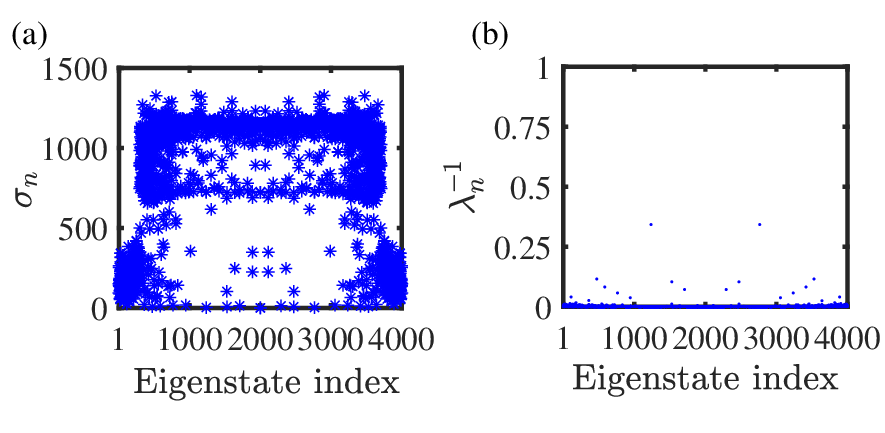}
\caption{(Color online)(a) $\sigma_n$ as a function of the eigenstate index. (b)The Lyapunov exponent $\lambda^{-1}_n$ as a function of the eigenstate index. Here, $v=0.2$, $\Delta=0.08$, $L=2000$, and $\phi=0$.}
\label{Fig11}
\end{figure}

For $v=0$ and $\Delta=0$, the system's energy spectra exhibit $E=\pm w$, leading to a substantial degeneracy at energy level $E$. This extensive degeneracy fosters the emergence of compact localized states within the flat bands, where the eigenstates are confined to a finite number of sites. However, with the introduction of quasi-periodic perturbations, the pristine all-flat-band structure is disrupted, revealing intricate localized properties. We choose $v=0.2$ and $\Delta=0.08$ for our discussion. Figures \ref{Fig11}(a) and \ref{Fig11}(b) show $\sigma_n$ and the Lyapunov exponent $\lambda_n^{-1}$ respectively, as functions of the eigenstate index. From Fig.~\ref{Fig11}(a), the values of $\sigma_n$ exhibit a large fluctuation, suggesting the existence of the critical states in such case. And as shown in Fig.~\ref{Fig11}(b), $\lambda_n^{-1} \to 0$, signifying the extended nature of the eigenstates. These findings collectively suggest that the system, under the influence of small $v$ and $\Delta$, resides in an extended but nonergodic phase.

\section*{Appendix C: The consistency between the reentrant localization and the TAI phase transition}
\setcounter{equation}{0}
\setcounter{figure}{0} \setcounter{table}{0} %
\renewcommand{\theequation}{C\arabic{equation}}
\renewcommand{\thefigure}{C\arabic{figure}}
\renewcommand{\thetable}{C\Roman{table}}

\begin{figure}[h]
	\includegraphics[width=0.5\textwidth]{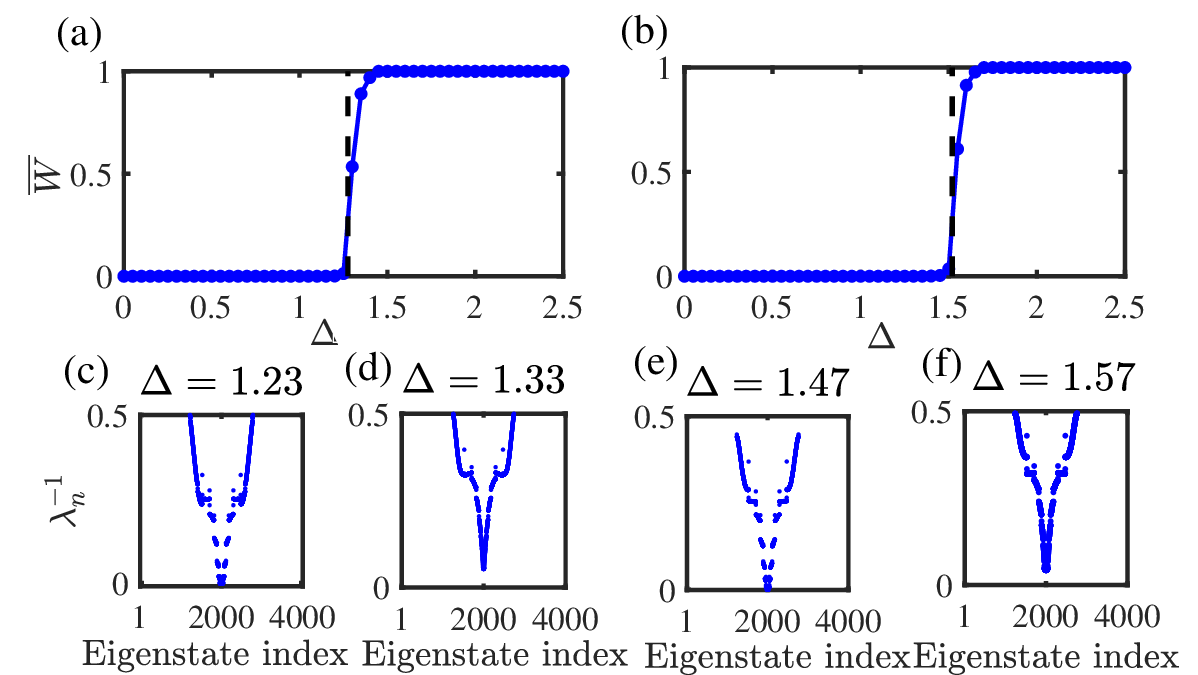}
	\caption{(Color online)The disorder-averaged winding number $\overline{W}$ as a function of the modulation strength $\Delta$ with $L=200$ for (a) $v=1.6$ and (b) $v=1.9$. In (a) and (b), the black dashed lines correspond to the TAI phase transition points. The distributions of the $\lambda_n^{-1}$ as a function of the eigenstate index with $L=2000$ for (c) $v=1.6$ and $\Delta=1.23$, (d) $v=1.6$ and $\Delta=1.33$, (e) $v=1.9$ and $\Delta=1.47$, and (f) $v=1.9$ and $\Delta=1.57$. All data in figures are averaged over $100$ disorder realizations.}
	\label{Fig12}
\end{figure}

To elucidate the correlation between reentrant localization and the TAI phase transition, we consider specific examples with $v = 1.6$ and $1.9$. The disorder-averaged winding number $\overline{W}$ is computed as a function of $\Delta$, as depicted in Figs.\ref{Fig12}(a) and \ref{Fig12}(b). For $v=1.6$ and $1.9$, the topological transition points are identified as $\Delta_c=1.28$ and $1.52$, respectively, determined by Eq.(\ref{eq8_2}). These transition points are highlighted by black lines in Figs. \ref{Fig12}(a) and \ref{Fig12}(b). It is evident that as $\Delta$ crosses the topological phase transition points $\Delta_c$, the value of $\overline{W}$ undergoes a jump from $0$ to $1$, signifying the emergence of the TAI. Simultaneously, we examine the corresponding Lyapunov exponents $\lambda_n^{-1}$ for all eigenstates near the topological phase transition points, illustrated in Figs.\ref{Fig12}(c)-\ref{Fig12}(f). For $\Delta$ values smaller than $\Delta_c$ [Figs.\ref{Fig12}(c) and \ref{Fig12}(e)], the Lyapunov exponents $\lambda_n^{-1}$ in the band-center region tend to $0$, indicative of the existence of delocalized states in the system. Conversely, when $\Delta$ surpasses $\Delta_c$ [Figs.\ref{Fig12}(d) and \ref{Fig12}(f)], all values of $\lambda_n^{-1}$ become greater than $0$, suggesting the system reenters the localized phase. Our findings suggest that the reentrant localization transition closely tracks the TAI phase transition.

\section*{Appendix D: Bott index for the topological phase transition}
\setcounter{equation}{0}
\setcounter{figure}{0} \setcounter{table}{0} %
\renewcommand{\theequation}{D\arabic{equation}}
\renewcommand{\thefigure}{D\arabic{figure}}
\renewcommand{\thetable}{D\Roman{table}}
\begin{figure}[h]
	\includegraphics[width=0.45\textwidth]{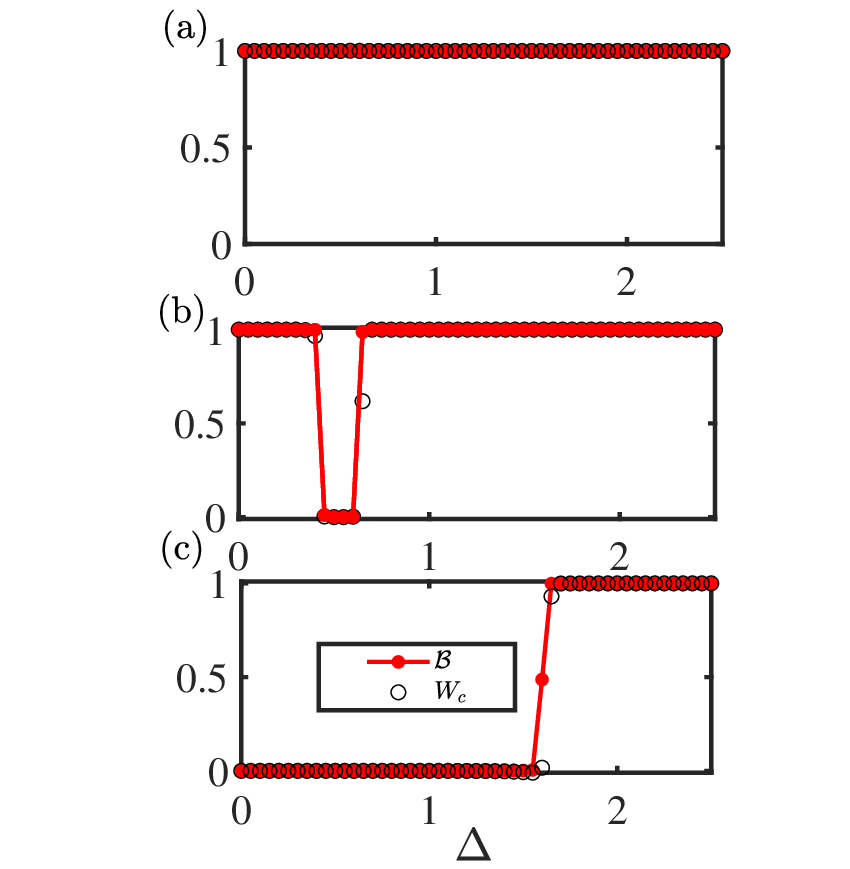}
	\caption{{(Color online)$\mathcal{B}$ and $W_c$ as a function of the modulation strength $\Delta$ under OBCs with (a) $v = 0.2$, (b)$v = 0.8$, and (c)$v=2$, respectively. Other parameters: $\phi=0$ and $L=500$.}}
	\label{Fig13}
\end{figure}
{Due to the absence of translation symmetry, it is natural to consider a real-space representation of topological invariants. One notable example is the Bott index $\mathcal{B}$. In the canonical representation, a Hamiltonian $H$ with chiral symmetry can be represented as follows structure\cite{LCH2021}:}
\begin{eqnarray}
H=\left[\begin{array}{cc}
0 & h \\
h^{\dag} & 0 \\
\end{array}
\right],
\label{D1}
\end{eqnarray}
{where $h$ is an $L_A \times L_B$ matrix, and $L_A$ and $L_B$ denote the number of sites in the $A$ and $B$ sublattices, respectively. Here, we set $L_A=L_B=L$. The homotopically equivalent flat-band version of the Hamiltonian $H$ is given by\cite{DZhang2020,LCH2021}:}
\begin{eqnarray}
Q'=\left[\begin{array}{cc}
0 & q \\
q^{-1} & 0 \\
\end{array}
\right],
\label{D2}
\end{eqnarray}
{where $q$ is a unitary matrix. The 1D Bott index($\mathcal{B}$) is then defined as\cite{LCH2021}:}
\begin{eqnarray}
\mathcal{B}=\frac{1}{2\pi i}{\rm{Tr\log}}(q^{-1}\tilde{\chi}q\tilde{\chi}^{-1}),
\label{D3}
\end{eqnarray}
{with $\tilde{\chi}$ being the position operator, expressed as $\tilde{\chi}={\rm{diag}}(e^{i\frac{2\pi}{L}1},e^{i\frac{2\pi}{L}2},...,e^{i\frac{2\pi}{L}L})$. The Bott index is particularly useful for disordered systems. In our numerical calculations, we set $\phi=0$ for simplicity. We focus on three example values of $v=0.2$, $0.8$ and $2$ to study the variation of the Bott index with the modulation strength $\Delta$, as shown in Figs.\ref{Fig13}(a)-(c). As a comparison, we also plot the winding number $W_c$ as the function of $\Delta$ in Fig.\ref{Fig13}.}

{For $v=0.2$, as $\Delta$ increases, both $\mathcal{B}$ and $W_c$ remain constant at unit, indicating that the system is in a topologically nontrivial phase, as shown in Fig.~\ref{Fig13}(a). For $v=0.8$, the system undergoes topological transitions between nontrivial, trivial, and again nontrivial phases as $\Delta$ increases. Correspondingly, both $\mathcal{B}$ and $W_c$ exhibit two sharp jumps between $0$ and $1$, shown in Fig.~\ref{Fig13}(b), which suggests the emergence of a RTAI. For $v=2$, increasing $\Delta$ causes both the Bott index and $W_c$ to jump from $0$ to $1$, indicating a TAI is induced by $\Delta$, as shown in Fig.~\ref{Fig13}(c). }

\begin{acknowledgments}
 Z.X. is supported by the NSFC (Grant No. 12375016), the Fundamental Research Program of Shanxi Province, China (Grant No. 20210302123442), and Beijing National Laboratory for Condensed Matter Physics (No. 2023BNLCMPKF001). Y. Zhang is supported by the National Natural Science Foundation of China (12074340). This work is also supported by NSF for Shanxi Province (Grant No. 1331KSC).
\end{acknowledgments}
\par
\par

\end{document}